\documentclass{article}

\usepackage[final]{neurips_2020}

\usepackage[utf8]{inputenc} % allow utf-8 input
\usepackage[T1]{fontenc}    % use 8-bit T1 fonts
\usepackage{hyperref}       % hyperlinks
\usepackage{url}            % simple URL typesetting
\usepackage{booktabs}       % professional-quality tables
\usepackage{amsfonts}       % blackboard math symbols
\usepackage{nicefrac}       % compact symbols for 1/2, etc.
\usepackage{microtype}      % microtypography

\usepackage{amsmath}
\usepackage{graphicx}
\usepackage{float}
\usepackage{subfig}
\usepackage{longtable}
\usepackage{booktabs}
\usepackage{multirow}
\usepackage{siunitx}
\usepackage{algorithm}
\usepackage{mathpazo}
\usepackage{mathtools}
\usepackage{tabularx}

\title{Soft Attention Convolutional Neural Networks for Rare Event Detection in Sequences}

\author{Mandar Kulkarni, Aria Abubakar \\ Schlumberger \\ (mkulkarni9, aabubakar) @ slb.com}

\begin{document}

\maketitle

\begin{abstract}

Automated event detection in the sequences is an important aspect of temporal data analytics. The events can be in the form of peaks, changes in data distribution, changes of spectral characteristics etc. 
In this work, we propose a Soft-Attention Convolutional Neural Network (CNN) based approach for rare event detection in sequences. For the purpose of demonstration, we experiment with well logs where we aim to detect events depicting the changes in the geological layers (a.k.a. well tops/markers).  Well logs (single or multivariate) are inputted to a soft attention CNN and a model is trained to locate the marker position. Attention mechanism enables the machine to relatively scale the relevant log features for the task. Experimental results show that our approach is able to locate the rare events with high precision.

\end{abstract}

\section{Introduction}

Automated event detection/localization in the sequences is an important aspect of temporal data analytics. In this paper, we propose a machine learning based approach for rare event detection in sequences and validate its efficacy for sequential well log data.

\begin{figure*} [!t]
\centering
\begin{tabular}{c c}

\includegraphics[width=180pt, height = 130pt]{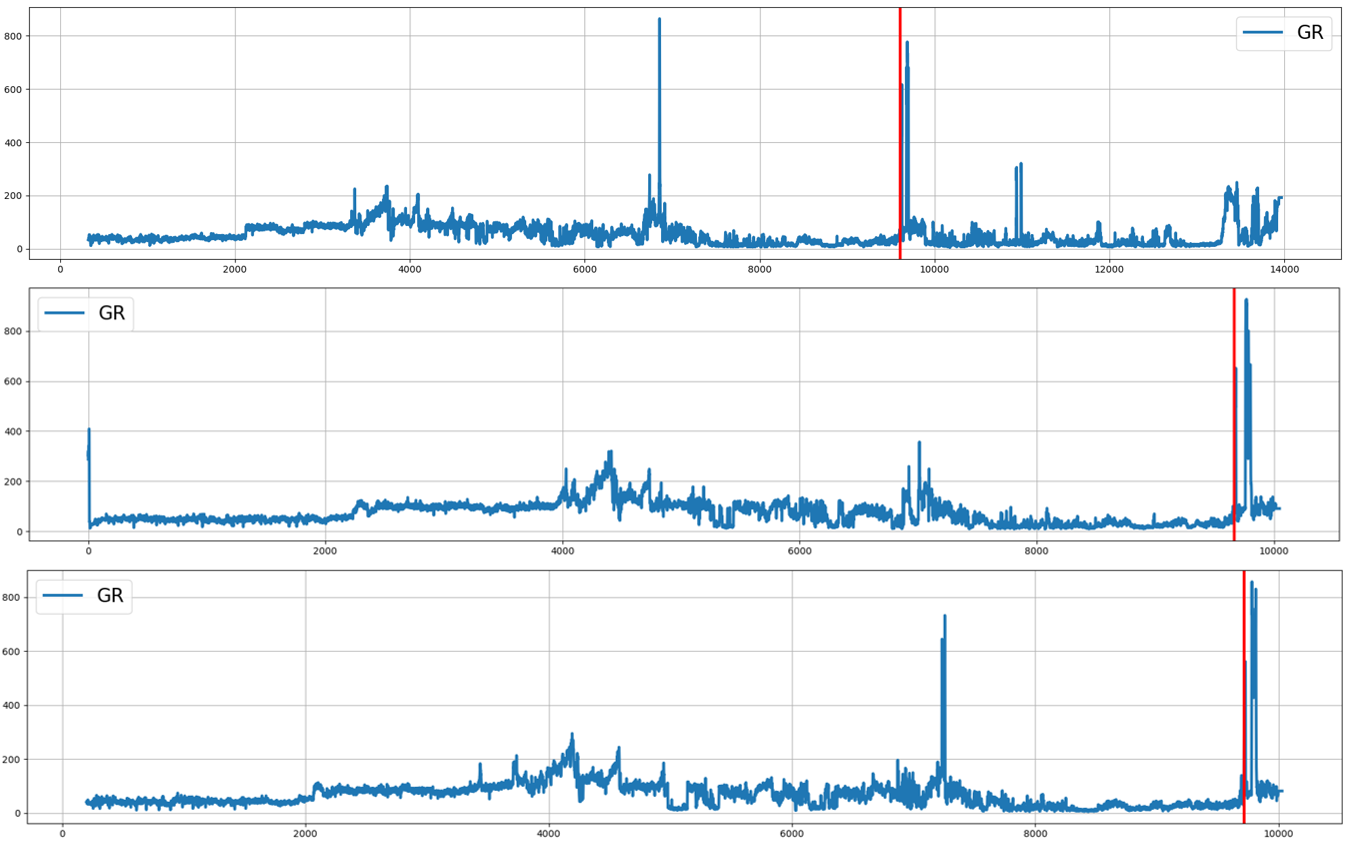}&
\includegraphics[width=180pt, height = 130pt]{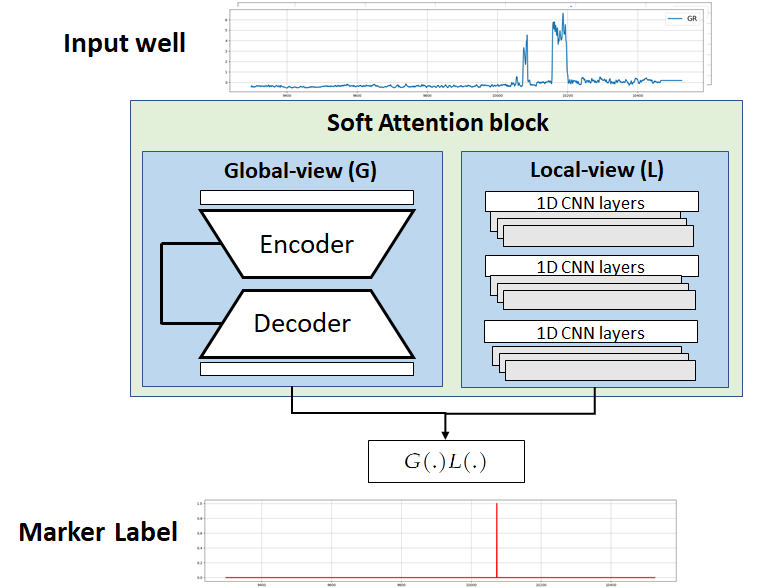}\\(a)&(b)\\

\end{tabular}
\caption{\label{fig:inpt1} (a) Sample well logs from Bakken dataset labeled with marker location (marked in red). x-axis indicates the depth (b) Proposed soft attention CNN architecture.}
\end{figure*}

Well logs are geological sequences which uniquely characterize sub-surface lithology. Typically, the measurements taken are Gamma Ray (GR), Resistivity (RES), Density (DEN) and the standard resolution of the measurements is 0.5ft.
The domain experts mark the depths in the well logs where there is a change in the geological layers. These labels are referred to as well tops or markers. Various markers are picked in all the wells in the basin in order to correlate the layer boundaries across the different wells. This problem is referred to as well log correlation. Fig. \ref{fig:inpt1}(a) shows sample well logs with GR measurements labeled with a marker location (indicated in red). Note that each well log has close to or more than 20k samples and only one of the index corresponds to the marker. This renders a marker localization a rare event detection task. 

We propose a Soft-Attention CNN framework for localizing these rare marker events in the well logs. For locating a particular marker over the log, all the input log patterns may not be relevant. Soft attention mechanism enables the machine to automatically adjust the scaling of the relevant log features for the marker picking. Experimental results show that the model is able to locate markers with high precision. Moreover, the same model architecture works well across the markers as well as with different log types, thus saving the compute effort on the case-specific hyper-parameter tuning. We observe that, during the manual top picking process, interpreters look at the entire well log to decide the relevant log section to focus on. We intuitively try to emulate the interpreter's workflow through soft attention.

\section{Methodology}

CNNs with soft attention (\cite{wang2017residual}, \cite{chen2016attention}) have shown promising results for image classification/segmentation problems. In this work, we demonstrate its application for the rare event detection in sequences. 

Fig. \ref{fig:inpt1}(b) depicts the proposed network architecture. 
Input to the model is the entire sequence (i.e. well log) and the network is trained to predict the marker location. For training the model, the location of an expert marker is converted to a one hot vector and used as the label. Under this setting, we train one model for each marker separately. Since the model is fully convolutional, it can be trained with input well logs of variable lengths.   

The well log is simultaneously input to Global-view (G) and Local-view (L) models. 
The global view model comprise of an encoder-decoder architecture. We resort to the U-Net (\cite{ronneberger2015u}) styled architecture where there are shortcut connections between the corresponding layers of the encoder and the decoder. The encoder model consist of 1D convolution layers followed by average pooling layers while the decoder model consist of 1D convolution layers followed by interpolation (up-sampling) layers. After each convolution layer, we use layer normalization and dropout layers. Due to the successive pooling operations in the encoder, the same convolution filter size provides a larger receptive field in the deeper layers of the encoder and thus helps to capture the log pattern variations over a much longer depth range. Hence, we refer to this model as the global-view model. The local-view model consists of a stack of 1D convolution layers. Since the convolution by nature is a local operation, it helps to model the log pattern variations over a shorter scale. 
 
The output tensor from the Global model is multiplied, element-wise, with the output tensor of the Local model as follows
\begin{eqnarray}\label{eq:q}
A(x) = G(x)L(x) 
\end{eqnarray}
 where $A$, $G$ and $L$ indicates the outputs of the attention block, the global-view model and the local-view model, respectively and $x$ indicates the well input. For $G$, we use 'tanh' activation, hence features from $L$ can be scaled positively/negatively as required. The scaled tensor $A$ is then mapped to the output layer using sigmoid activation.
 Fig. \ref{fig:inp} in appendix section \ref{sec:attn} shows the sample prediction result with attention scores and prediction probabilities for the marker.
 The convolution layers in global-view as well as local-view model consist of inception blocks (\cite{Szegedy_2015_CVPR}), where convolution is performed with multiple filter sizes and the feature maps are concatenated.  Experimentally, it is observed that, for the local-view model, dilated filters provide better results. In ablation study section, we experimentally validate the effectiveness of the global-local view model as compared to the results obtained with either model alone.

%  Intuitively, the global model looks at the larger well context and acts as an filter which p learn to scale the relevant log patterns at the output of local model. 

% for the global model output hence the range of values would be between -1 and 1. The layers in local model has Relu activation. The multiplied output is then mapped to label using sigmoid activation. 

In the numerical experiments, the labeled data are divided into training, validation, and test set. The model is trained end-to-end using binary cross entropy loss until the validation loss seize to improve. To mitigate the issue of sparse labels (i.e. label imbalance), we use a Gaussian label smoothing where a Gaussian kernel of a fixed scale is convolved with the one hot label. In the ablation study section, we demonstrate the effectiveness of label smoothing strategy.

\begin{figure*} [!t]
\centering
\begin{tabular}{c c}
\includegraphics[width=170pt, height = 100pt]{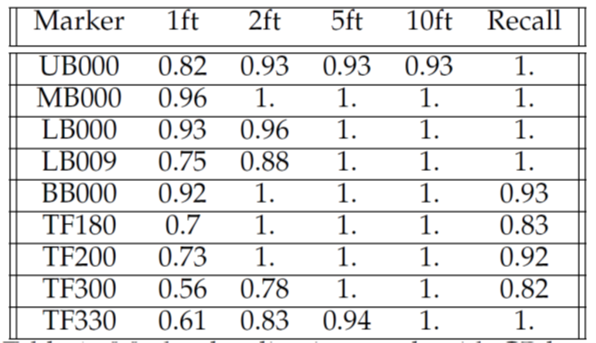}&
\includegraphics[width=170pt, height = 100pt]{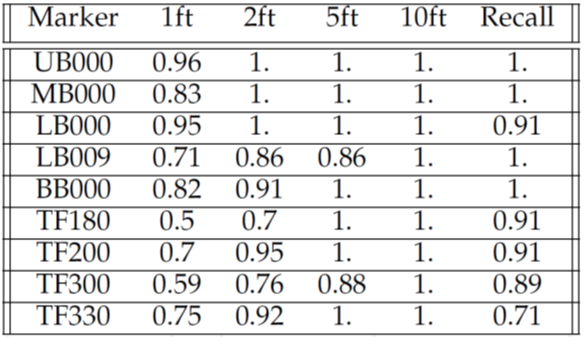}\\(a) GR & (b) GR-RES \\

\includegraphics[width=170pt, height = 100pt]{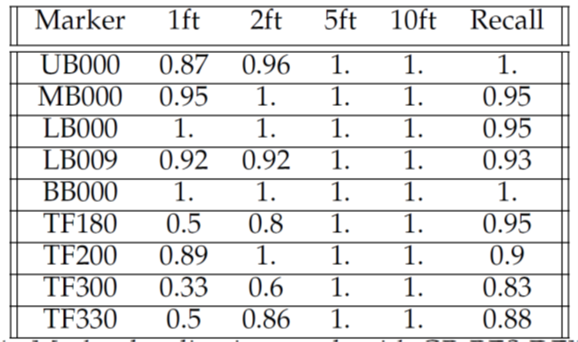}&
\includegraphics[width=200pt, height = 120pt]{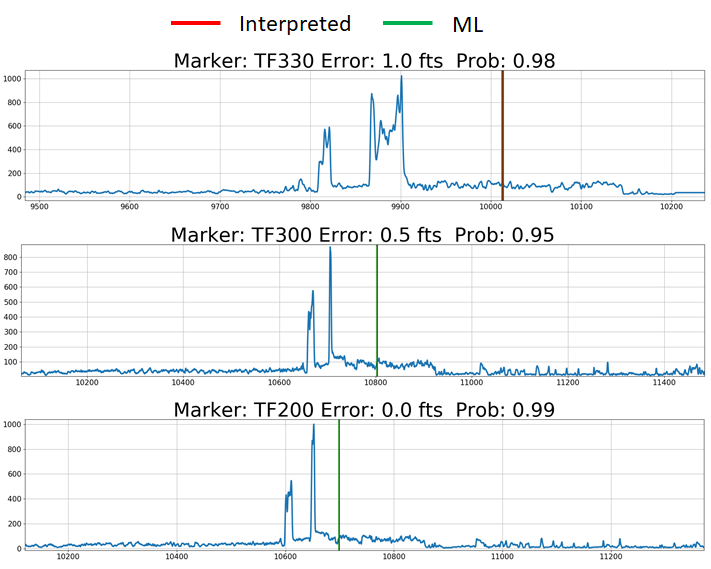}\\ (c) GR-RES-DEN & (d) Detection of subtle markers\\

\end{tabular}
\caption{\label{fig:inpt2} (a-c) Results with varied number of logs, (d) marker localization plots for subtle markers. Error indicates distance between Interpreted (Expert) and ML marker. }
\end{figure*}

\section{Evaluation metric}
To demonstrate the effectiveness of the proposed approach, we devised a following metric. For each marker type, the ML detection depth $D_{i}^{\text{ML}}$ and the error $e_i$ (in fts) between the ML marker and the manual marker is computed as follows
% \begin{subequations}
%   \begin{tabularx}{\textwidth}{Xp{0.5cm}X}
%   \begin{equation}
%     \label{eq-a}
%       e_i = |D_{i}^{\text{Expert}} - D_{i}^{\text{ML}}|
%   \end{equation}
%   & &
%   \begin{equation}
%     \label{eq-b}
%     D_{i}^{\text{ML}} = \arg\max_j P_{ij}
%   \end{equation}
%   \end{tabularx}
% \end{subequations}

\begin{subequations}
  \begin{tabularx}{\textwidth}{Xp{0.5cm}X}
   \begin{equation}
    \label{eq-b}
    D_{i}^{\text{ML}} = \arg\max_j P_{ij}
  \end{equation}
  & &
  \begin{equation}
    \label{eq-a}
      e_i = |D_{i}^{\text{Expert}} - D_{i}^{\text{ML}}|
  \end{equation}
  \end{tabularx}
\end{subequations}

$P_{ij}$ indicates the probability prediction over the depths. $j$ ranges over all the depths in the $i^{th}$ target well and $\max_j P_{ij}$ is the probability of the detection.

% where $e_i$ indicates the error, $D_{i}^{\text{Expert}}$ indicates the location of the expert pick while $D_{i}^{\text{ML}}$ indicates the ML prediction depth for the $i^{th}$ target well. The ML marker prediction depth is calculated as follows

% The marker predictions with the confidence score higher than the threshold are considered as valid detections. 

The precision and recall of the ML predictions is then in calculated as follows \begin{subequations}
  \begin{tabularx}{\textwidth}{Xp{0.5cm}X}
  \begin{equation}
    \label{eq-a}
      Prec_{d_T} = \frac{\sum_{i} I[e_i <= d_T]}{M}
  \end{equation}
  & &
  \begin{equation}
    \label{eq-b}
    Recall = \frac{M}{N}
  \end{equation}
  \end{tabularx}
\end{subequations}
Here, $I$ denotes the indicator function which evaluates to 1 if the condition is satisfied. $M$ denotes the number of ML pick detections. For each marker, we calculate the precision with multiple error tolerances where $d_T$ is set to 1ft, 2ft, 5ft, 10ft. For the recall, $N$ denotes the number of test wells.

To measure the performance on the dataset with all markers, we calculate the $F_1$-score as follows
\begin{eqnarray}\label{eq:q}
\text{$F_1$} = \frac{2 Mean (Prec_{d_T}) Mean (Recall)}{Mean (Prec_{d_T}) + Mean (Recall)}
\end{eqnarray}
where the mean is calculated across all the markers. For all the reported $F_1$ scores in this paper, $d_T$ is set to 2ft.

\subsubsection{Model uncertainty}

% For reliable performance of deep learning models in the real-world applications, its crucial to evaluate an uncertainty of the model prediction. 
To estimate the model uncertainty, we use Monte Carlo (MC) Dropout [\cite{gal2016dropout}] strategy. For each test well, we make multiple (30) predictions with dropout and the standard deviation of the detection depths is used as the proxy for the model uncertainty. 
The marker predictions with probability score more than a threshold and model uncertainty less than a threshold are considered valid detections. For all our experiments, we set the uncertainty threshold to 5 fts.

\section{Experimental results}
We now demonstrate the effectiveness of the proposed architecture under various scenarios. 
Note that the same network architecture was used for generating all the results.

\subsection{Description of the dataset}

For the purpose of experimentation, we use a public dataset collected from the Williston basin (Bakken). The wells have logs such as Gamma-Ray (GR), Resistivity (RES), and Density (DEN). We have a  dataset with 144 labeled wells. The subset of labeled wells with GR, GR-RES and GR-RES-DEN logs are 144, 122, and 118, respectively.
There are two variety of markers present in this dataset. 
The strong signature markers included in the study are UB000, MB000, LB000 and LB009. 
For these markers, a prominent signature is visible on the log. 
The subtle signature markers included in the study are BB000, TF180, TF200, TF300, and TF330. For these markers, subtle signature is observed on the log, which renders the detection challenging.

\subsection{Results under various scenarios}

Fig. \ref{fig:inpt2}(a) shows the marker localization result with GR log where 80\% of the wells are used for training and validation while 20\% wells are used for testing. It can be seen that the model is able to pick markers with high precision. Majority of the markers are detected within 5ft of error tolerance with relatively good recall.  
Fig. \ref{fig:inpt2}(b) and Fig. \ref{fig:inpt2}(c) shows the results with GR-RES and GR-RES-DEN logs with 80\% train-val wells. For the majority of markers, better precisions are obtained with multiple logs.

We further investigated the results for the challenging subtle markers. Fig. \ref{fig:inpt2}(d) shows the detection result for GR log with subtle markers. Interestingly, even for the markers that are barely visible on the log, our model is able to pick it accurately. We hypothesize that, though a local signature of the marker is not prominent, a global view of the well is guiding the marker localization. For visual validation of the accuracies, we plotted  error histograms for all the results. Plots are provided in appendix in section \ref{sec:eh}. 

We also performed ablation study to understand the effect of individual components. Results are provided in appendix in section \ref{sec:ab}. Table \ref{table:gl} in section \ref{sec:glc} also shows the F1-score comparison with an existing well correlation approach proposed in \cite{maniar2018machine}. Note that our approach provides relatively better results.

\section{Conclusion}

In this paper, we proposed soft attention convolutional neural network based approach for well log correlation. The combination of global-view and local-view model is proposed to locate the markers given a single or multiple input logs. The proposed method achieved high precision on marker picks where the same architecture reliably works well across the markers and variable number of logs. Importance of individual components is verified through Ablation study.

%\append{The source of the bibliography}
%\verbatiminput{example.bib}

%\twocolumn

\pagebreak

\bibliographystyle{seg}  % style file is seg.bst
\bibliography{example}

\newpage

\section{Appendix}

\subsection{Ablation study} \label{sec:ab}

In this section, we performed experiments to understand the effect of individual components on the final result. 

\subsubsection{Effectiveness of global-view and local-view model}\label{sec:glc}

To validate the effectiveness of the combination of global and local models, we performed an experiment, where we trained each model independently and compared their $F_1$ scores. Table \ref{table:gl} shows the $F_1$ scores comparison results for Bakken and Permian datasets. It can be seen that, in both the cases, the combined model outperforms the individual global and local model. Table \ref{table:gl} also shows comparison result with an existing well correlation approach.
%
% \begin{table}[h!]
% % \begin{center}
% \centering
% \begin{tabular}{||c c c ||} 
%  \hline
% Only Global (G) &  Only Local (L) & Global + Local \\  [0.5ex]
%  \hline\hline
%  0.915 & 0.929 & \textbf{0.937}\\ 
%  \hline
% \end{tabular}
% \caption{\label{table:gl} $F_1$ scores comparison for individual models.}
% % \end{center}
% \end{table}

\begin{table}[h!]
% \begin{center}
\centering
\begin{tabular}{||c c c c c||} 
 \hline
Dataset & Global (G)  &  Local (L) & Global + Local & \cite{maniar2018machine} \\  [0.5ex]
 \hline\hline
 Bakken & 0.915 & 0.929 & \textbf{0.937} & 0.91\\ 
 \hline
Permian & 0.913 & 0.821 & \textbf{0.939} & - \\ 
 \hline

\end{tabular}
\caption{\label{table:gl} Comparison of $F_1$ scores for individual models.}
% \end{center}
\end{table}

\subsubsection{Effectiveness of label smoothing}

To validate the effectiveness of Gaussian label smoothing, we performed an experiment without label smoothing. Table~\ref{table:ls} shows the $F_1$ scores comparison with and without label smoothing. It indicates that the label smoothing gives significantly better results.

%
% \begin{table}[h!]
% % \begin{center}
% \centering
% \begin{tabular}{||c c ||} 
%  \hline
% Without &  With \\  [0.5ex]
%  \hline\hline
%  0.549 & \textbf{0.937}\\ 
%  \hline
% \end{tabular}
% \caption{\label{table:ls}$F_1$ scores result for label smoothing.}
% % \end{center}
% \end{table}

\begin{table}[h!]
% \begin{center}
\centering
\begin{tabular}{||c c c||} 
 \hline
Dataset & Without &  With \\  [0.5ex]
 \hline
 Bakken & 0.549 & \textbf{0.937}\\ 
 \hline
Permian & 0.652 & \textbf{0.939}\\ 
 \hline

\end{tabular}
\caption{\label{table:ls} Comparison of $F_1$ scores for label smoothing.}
% \end{center}
\end{table}

\subsection{Error histograms}\label{sec:eh}
\begin{figure*} [!h]
\centering
\begin{tabular}{c c}

\includegraphics[width=170pt, height = 75pt]{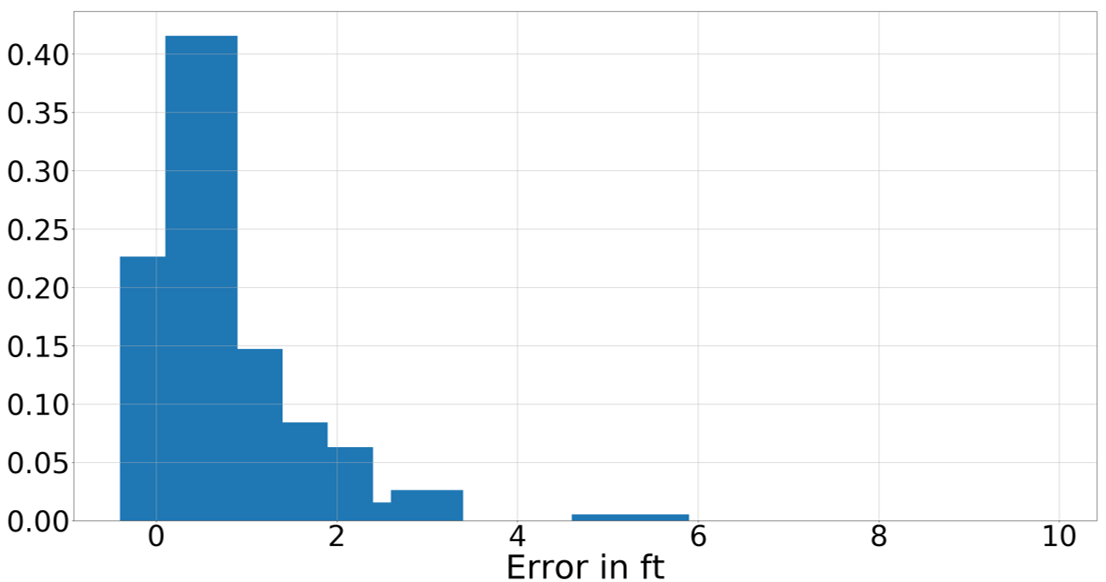}&
\includegraphics[width=170pt, height = 75pt]{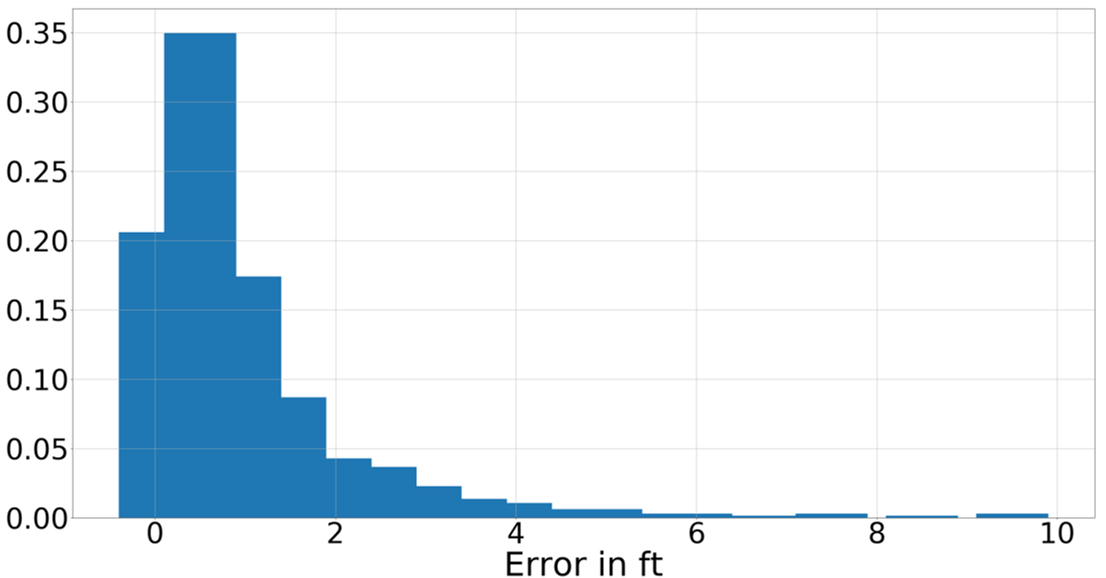}\\(a) GR (80\%) & (b) GR (25\%)\\

\includegraphics[width=170pt, height = 75pt]{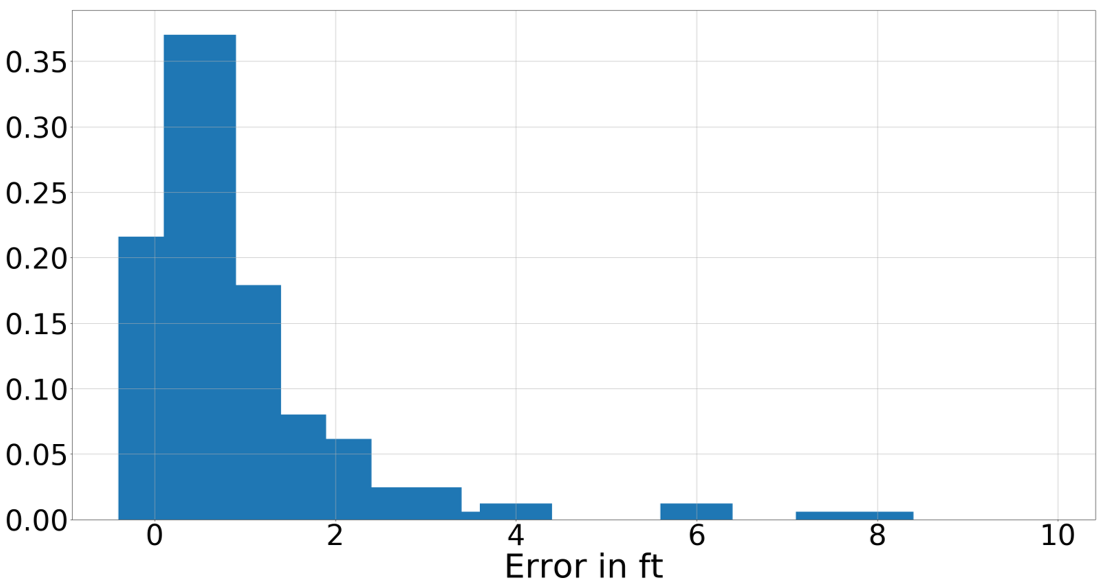}&
\includegraphics[width=170pt, height = 75pt]{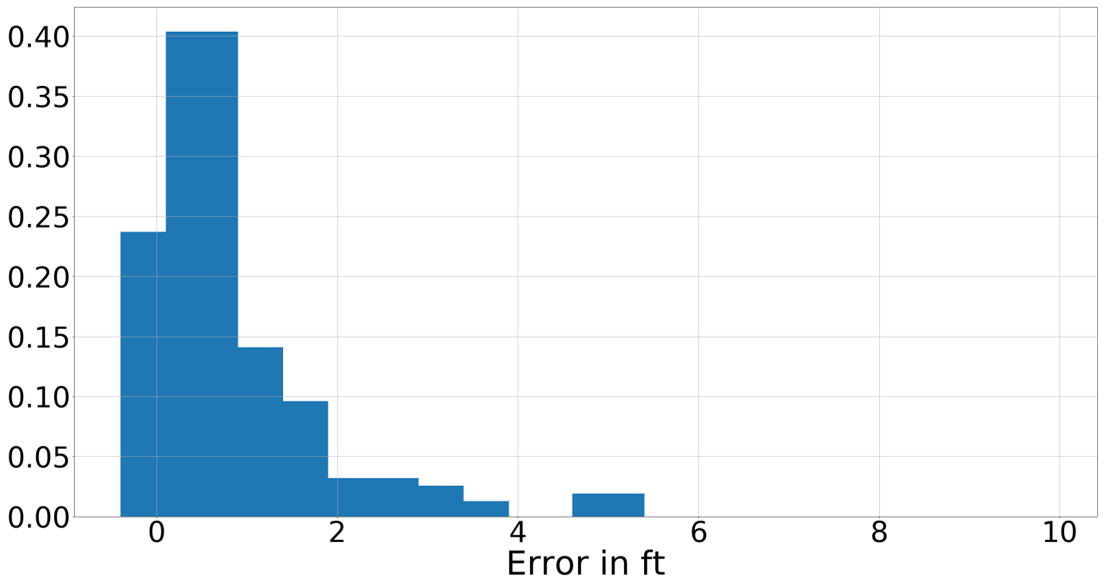}\\(c) GR-RES (80\%) &(d) GR-RES-DEN (80\%)\\

\end{tabular}
\caption{\label{fig:inpt4} Error histograms. x-axis indicates the error in fts. Percentage of wells used for training is indicated in the brackets.}
\end{figure*}

Fig. \ref{fig:inpt4} shows the histogram of marker errors. It provides the visual validation of high precision on the marker picks. Note that in all the cases, majority of the markers are picked within 2ft of the error tolerance. 

\subsection{Prediction result with attention scores} \label{sec:attn}

\begin{figure}
\includegraphics[width=\columnwidth]{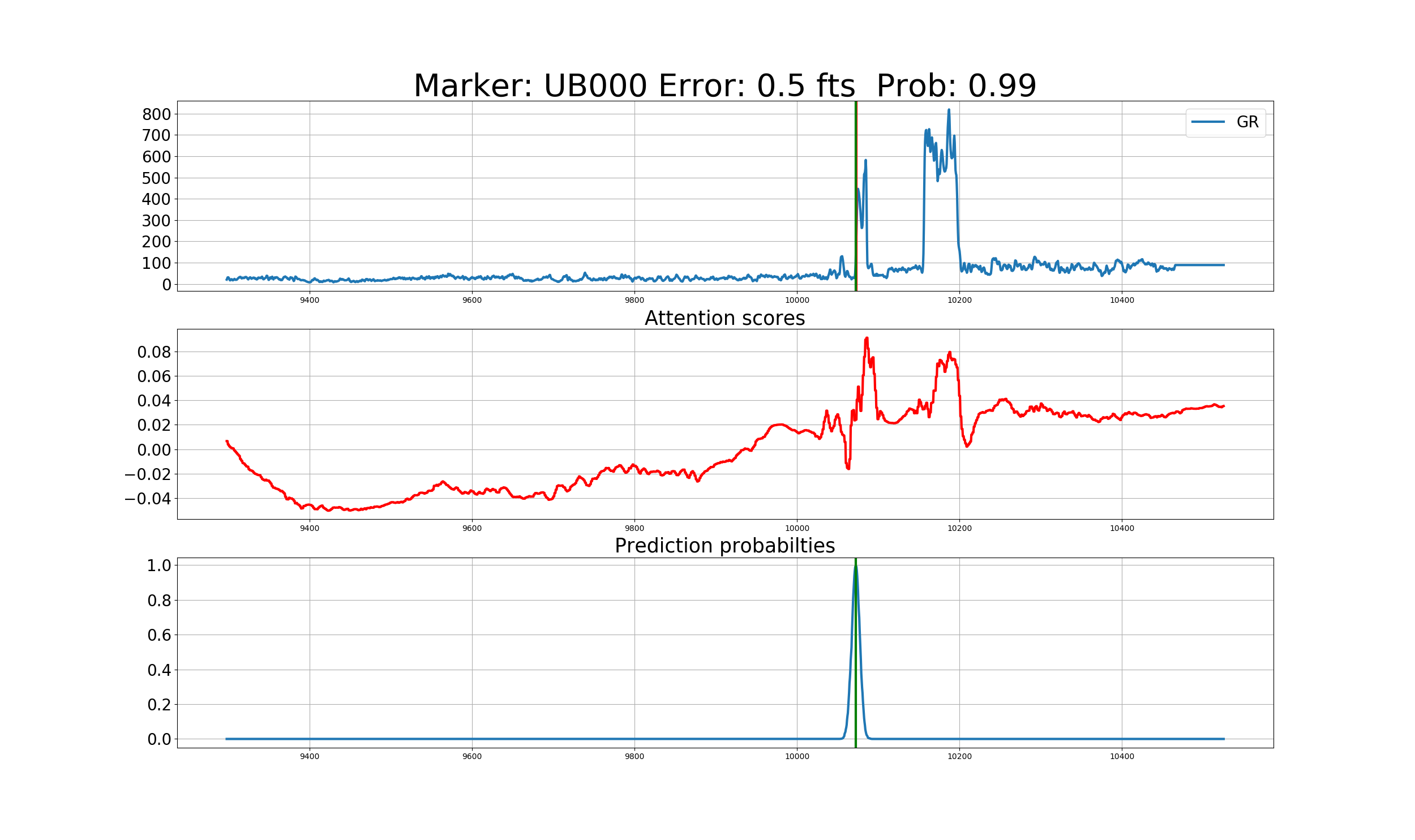}
\caption{\label{fig:inp} Prediction result with attention scores.}
\end{figure}

Fig. \ref{fig:inp} shows a marker prediction result along with the attention score over depths and prediction probabilities. Attention scores are obtained by averaging tensor $G$ along the feature axis. Note that the different depths have been assigned positive/negative attention score according to their relevance for the marker prediction. 

\end{document}